\documentclass[10pt,conference]{IEEEtran}
\newcommand{\sfy}[1]{} 

\renewcommand{\paragraph}[1]{ \textbf{#1} }

\usepackage{enumitem}
\usepackage{framed}

\usepackage{float}
\usepackage{wrapfig}

\usepackage{comment} %ability to begin/end large sections of comments.
\usepackage{url} % so we can use URLs in e.g.the bibliography
\usepackage{graphicx} % include graphics files
\usepackage{amsmath,amssymb,amsfonts} %extra math commands and symbols

\usepackage{tabularx}
\usepackage{mdframed}

\usepackage[backend=bibtex]{biblatex}

\title{Neverland: Lightweight Hardware Extensions for Enforcing Operating System Integrity}
\author{Salessawi Ferede Yitbarek, Todd Austin \\ University of Michigan}

\bibliography{main.bib}

\begin{document}

\maketitle

\begin{abstract}

The security of applications hinges on the trustworthiness of the operating system, as applications rely on the OS to protect code and data.
As a result, multiple protections for safeguarding the integrity of kernel code and data are being continuously proposed and deployed.
These existing protections, however, are far from ideal as they either provide partial protection, or require complex and high overhead hardware and software stacks.

In this work, we present Neverland: a low-overhead, hardware-assisted, memory protection scheme that safeguards the operating system from rootkits and kernel-mode malware.
Once the system is done booting, Neverland's hardware takes away the operating system's ability to overwrite certain configuration registers, as well as portions of its own physical address space that contain kernel code and security-critical data.
Furthermore, it prohibits the CPU from fetching privileged code from any memory region lying outside the \textit{physical addresses} assigned to the OS kernel and drivers (regardless of virtual page permissions).
This combination of protections makes it extremely hard for an attacker to tamper with the kernel or introduce new privileged code into the system -- even in the presence of kernel vulnerabilities. Our evaluations show that the extra hardware required to support these protections incurs minimal silicon and energy overheads.
Neverland enables operating systems to reduce their attack surface without having to rely on complex integrity monitoring software or hardware.

\end{abstract}

\section{Introduction}

Security mechanisms on a computer system heavily rely on the proper functioning of the OS kernel. 
If the kernel is compromised, other protections built on top of it can be bypassed.
\begin{comment}
\begin{figure}
	\centering
	\includegraphics[width=0.85\linewidth]{Chapters/figures/linux_vuln}
	\caption[Linux Kernel Vulnerabilities.]{\textbf{Linux Kernel Vulnerabilities.} New kernel vulnerabilities continue to be reported each year (data based on the National Vulnerability Database).}
	\label{figure-linux-vuln}
\end{figure}
\end{comment}

Today's systems commonly employ a secure boot mechanism to prevent the system from booting into a tampered kernel. However, once the boot process is completed, attackers could exploit software vulnerabilities to perform malicious actions such as overwriting kernel memory, executing malware in kernel-mode, or disabling driver integrity checks (example attacks can be seen in \cite{android-jop-root, pegasus, uroburos, motochopper, double-fetch-linux, double-fetch-win, mmap-escalation, dirty-cow}).
Completely eliminating vulnerabilities from today's large and complex kernels remains impractical.
For example, in 2017 alone, more than 200 Linux kernel vulnerabilities have been disclosed (based on the data from the National Vulnerability Database\footnote{https://nvd.nist.gov/. Vendor: ``linux'', Product:``linux\_kernel'', Keyword: ``linux kernel''}). 
Clearly, today's operating systems have a large attack surface as a result of their size and complexity. 
Furthermore, kernel modules (drivers) introduce additional attack surface, as they have full access to the kernel's address space.

\subsection*{Ongoing Challenges of Protecting the Kernel}
Recognizing their vast attack surface, a growing number of operating systems are deployed along with a continuous kernel integrity monitoring/enforcement mechanism.  Android distributions from major mobile vendors and recent versions of Microsoft Windows are two notable operating systems that have adopted this approach \cite{windows-7e, trust-issues, samsung-trustzone}.

These integrity checking mechanisms typically monitor kernel memory and CPU configuration registers to prevent malicious modifications.
Furthermore, they impose additional restrictions on the system, such as disallowing writable code pages in the kernel's address space.
Continuous integrity monitoring makes compromising the kernel quite challenging -- even in the presence of software vulnerabilities.

\paragraph{Guarding the Guardians.}
If integrity enforcement mechanisms run at the same privilege level as the kernel they are meant to protect, then vulnerabilities in the kernel could be exploited to subvert the integrity enforcement mechanism itself. 
For example, a kernel bug has previously been exploited to bypass Microsoft's Kernel Patch Protection (KPP) feature \cite{kpp-attack}. 

To shield these protection mechanisms from a compromised kernel, numerous previous works have proposed leveraging virtualization support available in today's systems \cite{garfinkel, vmwatcher, lares, hooksafe, secvisor, nickle, chen-noble}.
With this approach, the main OS runs on top of a hypervisor, while kernel protections are implemented in a hypervisor or another \textit{separate} virtualized operating system.
This provides additional protection as the hypervisor and the virtualized operating systems are isolated by the hardware. 
Microsoft Windows 10 has adopted this approach to protect its driver and kernel integrity enforcement mechanisms \cite{windows-7e}.

Alternatively, it is possible to leverage trusted execution environments, such as the TrustZone execution environment which is available on ARM CPUs.
Programs that run under TrustZone are isolated and protected from the rest of the system at the hardware level \cite{trustzone}.
Android devices commonly take advantage of TrustZone to isolate the kernel integrity mechanisms from the main kernel  \cite{samsung-trustzone, trust-issues}.

Even with the extra hardware-level isolation, software vulnerabilities in the kernel integrity enforcement mechanisms themselves remain exploitable. 
For instance, attacks against multiple generations of TrustZone-based kernel protections have been shown \cite{qsee-cve,trust-issues}.
In addition, virtualization-based security relies on hypervisors, which themselves have a large attack surface -- as proven by serious vulnerabilities that continue to be discovered in hypervisors \cite{hyperv-rce-2, hyperv-rce, hyperv-prev-esc, hyperv-prev-esc-2}.
Furthermore, running operating systems on top of a restrictive hypervisor incurs performance overheads \cite{vbs-analysis, hypernel}

\subsection*{Hardware Mechanisms for Safeguarding Kernel Integrity}

In this work, we present an efficient hardware-based protection, dubbed Neverland, which can be used to harden the operating system against kernel-mode malware and rootkits.  Neverland does not rely on a complex (and potentially vulnerable) kernel integrity monitoring software and incurs zero runtime performance overhead.

Once the kernel boots, Neverland's hardware irreversibly takes away the operating system's ability to overwrite certain configuration registers and portions of the physical address space. 
More specifically, we ``lock'' the kernel's code and read-only sections, security-critical configurations (e.g., driver signing configuration flags, system call table), and CPU configuration registers that store kernel entry points. 
These locked registers and memory regions can only be modified after rebooting the system.
Neverland also prohibits the CPU from fetching kernel-mode code from any memory region lying outside the physical addresses assigned to the OS kernel and drivers (regardless of virtual page permissions). 

Note that the code, data, and registers mentioned above are guarded or periodically scanned by a typical kernel integrity monitor. Locking these components at the hardware-level obviates the need to continuously run an integrity monitoring software.

We use a  \textbf{hardware permission table} to mark portions of the physical address space as read-only, executable, privileged, and locked. 
On every instruction fetch and load/store, the CPU checks the permission table to determine the legality of the memory operations.

A number of embedded CPU architectures specify a form of (optional) physical memory permission tables (e.g., ARM MPU and RISC-V PMP specifications). 
These ISA extensions are normally used to implement a lightweight memory protection scheme in low-resource microcontrollers that do not have virtual memory support \cite{zele-sfi, privilege-overlays}. 
CPUs with virtual memory support do not generally implement these physical memory permission schemes.
The hardware support required by Neverland is a variation of such permission tables.
We leverage immutable hardware tables to provide low-overhead operating system protections -- without requiring complex integrity monitoring software that rely on hypervisors or trusted execution environments. 
In addition, we also introduce the notion of ``locked'' CPU configuration registers to prevent interrupt and system call hijacking.

Maintaining a separate physical address permission table has advantages over existing hardware-level privilege escalation defenses such as Intel Supervisor Mode Execution and Access Prevention (SMEP and SMAP) and ARM Privileged-eXecute-Never (PXN). 
SMEP/SMAP and PXN are under the total control of the operating system and can be disabled if the kernel is compromised. 
Neverland side-steps this issue by stripping away the operating system's ability to overwrite the permission table entries once the boot process is complete.
Furthermore, since the physical address permission tables do not rely on virtual page permissions, attacks that manipulate the page table entries \cite{motochopper} are prevented.

We validated the efficacy of these protections by prototyping the hardware extensions in Spike (the official RISC-V emulator) and running a minimally modified version of the Linux kernel on the emulator. 
Our evaluations based on RTL simulation and synthesis show that the hardware extensions required by Neverland incur minimal silicon and energy cost, and that integrating the proposed protections does not incur any performance overhead beyond the 10s of milliseconds that are required to setup permissions at boot time.

In summary, we make the following contributions towards safeguarding the operating system's memory (code/data) from kernel-mode malware and rootkits:

\begin{itemize}

	\item We present a hardware-based memory protection scheme that can protect an operating system from kernel-mode malware and rootkits. The protection works by stripping away some of the kernel's powers once the boot process is done. We show how this protection can be applied without incurring any runtime performance overhead.  
	
	\item We demonstrate the practicability of the approach by integrating the proposed protections into an emulated RISC-V CPU, and by making minimal changes to the Linux kernel to take advantage of these hardware protections.  We also show the hardware and energy overheads of the additional hardware (a permission table) are minimal.
\end{itemize}

\sfy{Mondrian Memory Protection}

\section{Operating System Kernel Integrity}
\label{sec-os-background}
\sfy{mention iOS inegrity checks?}

The security and trustworthiness of applications depend on protections provided by the operating system.
A malware that is able to run at the same privilege level as the OS kernel 
can gain full control of a system and even render malicious activities invisible to the rest of the system.
As a result, software defenses or forensics techniques cannot reliably operate on top of a compromised OS.
Numerous protections are continuously being introduced to harden operating systems against attacks. 
These mitigations have made it significantly harder to compromise modern operating systems.

This section provides an overview of the attack vectors exploited by kernel-mode malware and rootkits, and describes existing kernel protections and their limitations. 

\subsection{Kernel-Mode Execution and Rootkits}
\label{sec-kernel-attacks}
The most severe compromises on an operating system would result in running arbitrary malicious code in kernel-mode (i.e., code that runs with kernel privileges). 
Some types of kernel-level malware, referred to as rootkits can even hide their existence and remain undetectable on the system \cite{subverting-windows}.
%\footnote{The term rootkit is sometimes used to describe non-malicious kernel or root code -- such as a security monitor. In this dissertation, we use the word rootkit to refer to malicious and stealthy kernel-mode malware.}, 

Despite the increasing sophistication and volume of attacks on the operating systems, kernel-mode malware and rootkits need to fundamentally rely on one of these three techniques to achieve privileged execution:

\begin{itemize}[leftmargin=*]
	
\item \textbf{Installing Malicious Drivers:} This is one of the oldest and most straightforward ways to run a malicious privileged code. Drivers/kernel modules can potentially tamper with any data in the system since they run at the same privilege level as the kernel. Modern operating systems make malicious driver installations difficult by requiring drivers to be signed by a trusted key \cite{macos-system-integrity, windows}.

\item \textbf{Privilege Escalation Attacks:} Attackers have repeatedly managed to discover mechanisms to leak or overwrite kernel memory.
These mechanisms range from exploiting memory safety and integer overflow/underflow errors in kernel-mode code \cite{motochopper, mmap-escalation, pegasus}, to exploiting race conditions \cite{dirty-cow, double-fetch-win, double-fetch-linux}, to abusing hardware bugs \cite{rowhammer-escalation}. Such exploits are convenient for injecting code into kernel memory or for executing user space code at an escalated privilege \cite{android-handbook}. 
In addition to exploits that function by corrupting or leaking kernel memory, it has been shown that nearly all major operating systems had the same bug in their system call handlers that enabled directly running user space code in kernel-mode \cite{sysret-cve}.
Evidently, there are numerous avenues for privilege escalation attacks.

Privilege escalation attacks are not mitigated by driver signature checks as such attacks can inject code into kernel space without passing through the driver loader logic. Furthermore, privilege escalation bugs which allow overwriting portions of kernel memory can themselves be used to disable code signature checks \cite{pegasus}. 

Operating systems are too large and complex to completely eliminate vulnerabilities that can lead to privilege escalation attacks.
As we will describe in the next section, modern CPUs enable operating systems to restrict which code pages can be executed in privileged mode. This hardware feature makes privilege escalation attacks more challenging even in the presence of certain vulnerabilities. 

\item \textbf{Code Patching:} It is also possible for attackers to modify kernel code to redirect execution or bypass certain security checks \cite{subverting-windows}. Since operating systems typically mark their code as read-only once loaded into memory, such type of code patching attacks need to rely on a bypass that enables writes to protected code pages or files.
\end{itemize}

\paragraph{Rootkit Stealth Techniques.}
Once attackers introduce malicious code into the system, they can employ additional stealth techniques to hide their existence. This is achieved by making network activities, processes, and files invisible to user space programs. Techniques for achieving stealth can be grouped into three categories \cite{subverting-windows}:
\begin{itemize}[leftmargin=*]
	\item  \textbf{System call and interrupt descriptor table hooking:}
	On a system call, the kernel consults the system call table (also referred to as System Service Descriptor Table in Windows ) to locate the kernel function that corresponds to a system call.
	On some CPU architectures, both interrupts and system calls are handled using a single dispatcher code.
	On the other hand, on x86 CPUs, the kernel and the CPU rely on the interrupt descriptor table (IDT) to execute the kernel code that corresponds to a specific interrupt request. 
	System call tables and IDTs are essentially a list of function addresses. 
	Rootkits can overwrite entries in these tables to execute malicious code on a system call or interrupt \cite{subverting-windows}.
	By hijacking system calls and interrupts, attackers can control what values the kernel returns to user space programs. 
	For example, a rootkit can hide a file or process by removing it from a list that a system call returns to a user space program.
	
	CPU architectures typically have a dedicated register that stores addresses of kernel entry points for system call and interrupt handling (e.g., on RISC-V, the stvec register typically holds the address of the kernel's system call/interrupt handler, while the IDTR register holds the address of the IDT in x86). Since direct modification of system call tables or the IDT might trigger certain rootkit detectors, some rootkits create a new IDT or system call handler and overwrite the address in these special purpose registers to point to the new addresses \cite{subverting-windows}.
 \item \textbf{Hooking code pointers:} It is relatively easier to detect changes to pointers in the system call or interrupt descriptor tables. As a result, advanced rootkits try to redirect execution by hijacking dynamic code pointers in the kernel. Such type of stealthy hooking is preferred by today's advanced rootkits \cite{hookscout} as 
 detecting malicious changes to an arbitrary code pointer is a more challenging task.
 
 \item \textbf{Direct Kernel Object Manipulation (DKOM):} A kernel typically has various well-documented data structures to track system information such as running processes/threads and list of open network connections. Rootkits can manipulate these data structures directly to make their execution and network footprint invisible \cite{subverting-windows}.
\end{itemize}

\subsection{Shielding the Operating System from Attacks}
\label{background_defenses}
Due to the prevalence and severity of attacks that target operating systems, numerous techniques for protecting the kernel have been extensively studied and deployed. In this section, we will highlight these protections, and motivate the necessity of the hardware extensions we propose in the next section.

\paragraph{Secure Boot.}
\label{sec-secure-boot}
Running a trustworthy operating system requires that the system initially boots into an untampered kernel.  
If the BIOS loads a compromised kernel, or if the kernel loads a compromised driver, then all other protections built into the operating system can be rendered ineffective.

Secure boot refers to a mechanism that verifies the integrity of the BIOS and kernel to ensure the system boots into an untampered state.
This is achieved by deploying a digitally signed or encrypted kernel and BIOS code, which the bootloader can later verify using the appropriate encryption key. 
It is aimed at detecting modifications made prior to booting, but provides no guarantees the operating system will remain untampered once it starts running.

The protections we discuss below, and the protections we propose in the next section,  generally assume that a secure boot mechanism ensures the system initially boots into an untampered kernel.
That being said, secure boot implementations are certainly \textit{not} impenetrable and can be vulnerable to certain advanced attacks \cite{secureboot-attacks}.

\paragraph{Driver Signing.}
\label{sec-driver-signing}
Operating systems rely on driver signing to prevent malicious drivers from being loaded into the system. 
Recent versions of Microsoft Windows and macOS only allow signed drivers by default \cite{windows-7e, macos-system-integrity}.
The Linux kernel also supports kernel module signature verification. 

Even if the widespread adoption of driver signing has made it significantly harder to install malicious drivers, attackers have repeatedly found ways to bypass this protection.
Attacks that disable code signing by exploiting a bug in the kernel or a legitimate driver have been discovered in the wild \cite{uroburos, pegasus}
 
Furthermore, driver signing mechanisms cannot defend against privilege escalation attacks (discussed above) since those attacks avoid going through the driver loading mechanism altogether and do not trigger any driver signature checks. 

%\sfy{cite user to kernel mode privilege escalation attacks - on ios for example}
%Recent research has also shown that major operating systems were vulnerable to a privilege escalation bug that resulted from mishandling of 

\paragraph{Supervisor Mode Access/Execution Prevention.}
\label{sec-smap}
In a typical virtual memory system, a bit in the page table entries (PTEs) marks a page as user-mode or kernel-mode page.
The CPU can use this bit to prevent user space programs from accessing kernel memory.
  
Modern CPUs provide another layer of page-based protection to prevent arbitrary user space code from executing with escalated privileges.
These CPUs check the user/supervisor bit in the PTEs, and prevent code from a user-mode page from being executed in kernel-mode.
Intel's Supervisor Mode Execution Prevention (SMEP) \cite{intel-sdm-3a} and ARM's Privileged-Execute-Never (PXN) \cite{arm-guide} are two examples of this technology.
 
Intel Supervisor Mode Access Prevention \cite{intel-sdm-3a}, another related protection, uses the user/supervisor bit in PTEs to prevent the kernel from directly accessing user space memory. 
This prevents the operating system from erroneously dereferencing a pointer to a user space memory (which could potentially contain malicious instructions or data set up by a user space program).

These CPU extensions make privilege escalation attacks quite challenging even in the presence of kernel bugs. 
These protections, however, are under the full control of the operating system itself. 
As a result, an oversight or bug in the kernel could be exploited to disable these protections.
Kemerlis et al., for example, have demonstrated how the fact that the Linux kernel ``mirrors'' the entire physical memory in its own address space can be abused to bypass SMEP and PXN \cite{ret2dir}. 
Memory safety bugs have also been exploited to circumvent PXN and tamper with kernel data structures on Android \cite{android-jop-root}.

\sfy{how about pxn}
More importantly, SMAP and SMEP can be disabled by clearing a single bit in x86's CR4 control register. As a result, code-reuse attacks can be used to disable SMAP/SMEP by modifying the value of the CR4 register \cite{smep-xen-attack,smep-rop}.
The additional kernel integrity monitoring layers in Windows (discussed below) attempt to defend against this bypass technique by monitoring the state of the CR4 register \cite{windows-7e}.

%As we detail in the next section, Neverland sidesteps these issues since it a) avoids relying on the virtual memory system for providing protections, and b) cannot be disabled even by privileged code, once it is enabled at boot time.

\begin{table*}[t]

\begin{centering}

\caption[Existing OS Kernel Protections and their Limitations/Drawbacks]{\textbf{Existing OS Kernel Protections and their Limitations/Drawbacks.}}
	
\begin{tabularx}{\textwidth}{|X|X|X|}

\hline
\textbf{Protection} (\S \ref{sec-driver-signing})                                  & \textbf{Known Limitations or Drawbacks}                                                                                                                                                                                                   \\ 
\hline
Secure Boot               & Can be undermined by vulnerabilities in bootloader; provides no guarantees the operating system will remain untampered once it starts running. \\
\hline
Driver/Kernel Module Signing                  & Can be disabled by exploiting OS or driver vulnerability; cannot catch privilege escalation attacks that do not directly load drivers                                                                                                     \\ 
\hline
Supervisor Mode Execution/Access Prevention and Privileged Execute Never  & Run under the full control of the untrusted operating system                                                                                                                                                                                                                                       \\ 
\hline
In-Kernel Integrity Monitoring               & Runs at the same privilege level as the untrusted kernel                                                                                                                                                                                 \\ 
\hline
Hardware-Isolated Integrity Monitoring      & Requires complex and high overhead software stacks (such as a hypervisor and/or a separate trusted kernel), which are still prone to exploits in the presence of software bugs.                                                           \\ 
\hline
Hardware-Level Kernel Memory \mbox{Monitoring}        & Implementations may require significant amount of new hardware, periodic full memory scans, and periodic cache flushes. It is also challenging to identify the semantic meaning of raw bytes in RAM without having access to more information about the system's state.  \\
\hline
\end{tabularx}

\end{centering}

\label{table-prot-compare}

\end{table*}

% & \textbf{Objective}                                                                                        
% & Ensure the system boots into an untampered OS.                                                            
% & Prevent execution of malicious drivers                                                                    
% & Prevent the kernel from being ``tricked'' to directly execute or access code/data from userspace memory   
% & Prevent malicious modification of kernel code and security critical data                                  
% & Run integrity enforcement software shielded from exploits on the main kernel.                             
% & Monitor the integrity of the kernel at the hardware-level, being as independent as possible from the main OS

\paragraph{Kernel Integrity Monitoring and Enforcement.}
\label{sec-kernel-integrity}
%As highlighted above, an attacker can run malicious code by either disabling driver signature checks, or by relying on various privilege escalation bugs.
%Even in cases where the attacker may not have achieved privileged execution, vulnerabilities in the operating system could be exploited to modify kernel memory.
Eliminating all potential vulnerabilities from an operating system remains to be an elusive goal.
Therefore, to safeguard the kernel's integrity even in the presence of vulnerabilities, certain operating systems (notably Microsoft Windows and some Android distributions) provide an additional security layer for monitoring/enforcing kernel integrity \cite{samsung-trustzone, windows-7e}. 

64-bit versions of Microsoft Windows are protected by PatchGuard -- a Kernel Patch Protection\footnote{Part of the original motivation for Windows' kernel patch protection feature was improving system stability by preventing unsafe changes to the kernel even by non-malicious third-party programs, such device driver and anti-virus software.} mechanism that is directly integrated into the kernel. PatchGuard periodically scans kernel code, CPU control registers, and data structures that are commonly targeted by malware and rootkits (e.g., interrupt descriptor tables, system service descriptor tables, the kernel stack, and kernel configuration flags) \cite{kpp-attack, windows-7e}.

Unfortunately, kernel integrity protections that run at the same privilege level as the kernel can be disabled by exploiting a vulnerability in the kernel itself.
As a result, PatchGuard has been subject to attacks that relied on kernel exploits \cite{kpp-attack}.
More secure implementations run the integrity protection module in a hardware-isolated environment -- such as a separate virtual machine or a TrustZone secure world (discussed below). Windows 10 and Android distributions used by major vendors are examples of commercial operating systems that have adopted this approach.

\paragraph{Virtualization-Based Kernel Protection.}
Virtualization-based kernel security solutions deploy the kernel in a virtual machine (VM) running on top of a hypervisor.  
Since the hypervisor has full control over VMs running on top of it, powerful defenses can be implemented inside the hypervisor or inside a separate VM.
Defenses implemented in this manner have an advantage over defenses built directly into the kernel as the hypervisor and the different VMs are fully isolated by the hardware -- making them harder to attack.

Previous works have demonstrated how hypervisors can be used to securely monitor activities in the guest OS \cite{garfinkel, vmwatcher, lares}, to detect code pointer hooking in the kernel \cite{hooksafe}, and prevent malicious privileged code execution \cite{secvisor, nickle}.

Windows 10 has also recently introduced a virtualization-based kernel protection scheme \cite{windows-7e}.
This protection enables security-critical tasks such as driver integrity verification, kernel integrity enforcement, and credential management to run in a separate virtual machine that is fully isolated from the Windows operating system.
This isolation protects security-critical processes from being compromised by a privileged malicious code.  

One challenge of virtualization-based security is that the hypervisor itself has a non-trivial attack surface \cite{samsung-trustzone}. For instance, multiple remote code execution and privilege escalation vulnerabilities have been discovered in Microsoft  Hyper-V in 2018 alone \cite{hyperv-prev-esc, hyperv-rce, hyperv-rce-2}. 
To counter this issue, additional protections for defending the hypervisor itself have been proposed \cite{hypersafe, nohype, hypersentry}.
A tiny hypervisor that is amenable to manual audits and verification has also been presented in \cite{secvisor}. %and formally verification has also been proposed. 
While these are all promising approaches, creating a fully verified commercial hypervisor that is bug-free remains to be a hard problem.

Aside from the security issues, running an OS in a virtualized environment also incurs performance overheads of up to 30\%, which is undesirable for certain deployments \cite{hypernel}. 
Furthermore, a study on the Microsoft Windows' virtualization-based protection has shown that system calls from untrusted applications can be up to 200x (2000\%) slower as a result of page table entry modifications the hypervisor needs to make on each system call \cite{vbs-analysis}.

\paragraph{TrustZone-Isolated Protection.}
Some ARM CPUs implement TrustZone, a hardware extension that enables isolating software and hardware components into secure and non-secure worlds \cite{trustzone}.
The hardware prevents the non-secure world from accessing the resources of the secure world.

Ge et al. proposed using this hardware feature for protecting kernel integrity monitors \cite{sprobes}. 
Android distributions deployed by major vendors use this approach to isolate the kernel integrity enforcement mechanisms from the main kernel \cite{samsung-trustzone, trust-issues}. 
They leverage TrustZone's capabilities by running the main operating system (i.e., Android) in the non-secure world, while running another lightweight (trusted) kernel in the secure world. 
Security-critical tasks such as credential management and integrity checking run on top of the trusted kernel, shielded from the main operating system.

Protections based on TrustZone, or similar trusted execution environment (TEE) are preferred over the virtualization-based approach as TEE-based protections do not need to rely on a large and complex hypervisor for isolation \cite{samsung-trustzone}.  
Nevertheless, TEEs cannot fully thwart attacks against a vulnerable code running in the hardware isolated secure-world, and attacks have been demonstrated against multiple generations of mobile TEEs \cite{rosenberg2014qsee, trust-issues, qsee-cve,qsee-writeup}.
In fact, trusted kernels commercially deployed in TEEs still lag behind mainstream operating systems in terms of built-in exploit mitigations \cite{trust-issues}.

\paragraph{Hardware Mechanisms for Kernel Integrity.}
\label{sec-hardware-monitor}
A number of previous works have proposed relying on hardware extensions or peripheral devices for monitoring the integrity of the kernel \cite{grim, hypersentry, vigilare, mguard, kimon, copilot}.

One category of the hardware-level defenses involves using a hardware peripheral, typically attached to the PCI bus, to take a snapshot of the physical memory \cite{grim, copilot, hypersentry}.
The snapshot is then analyzed to verify the kernel's integrity.
The hardware peripheral could be a custom hardware \cite{copilot}, a GPU \cite{grim}, or a network interface card \cite{hypersentry}

Repeatedly taking full memory snapshots can be slow and expensive, especially on systems that have a large memory. Furthermore, transient memory modifications that happen in between snapshot intervals can be missed \cite{vigilare}.
There is another class of protections that avoid periodically taking full memory snapshots, and instead monitor the memory bus \cite{kimon, mguard, vigilare}.
However, these bus snooping defenses require a significant amount of new hardware, including a standalone microcontroller that has its own private memory, DRAM controller, and DMA engine. 
In addition, bus monitoring hardware that only snoop the DRAM bus require periodic cache flushes to ensure they see all changes to memory \cite{mguard}.

These hardware mechanisms are designed to be minimally invasive by operating independent of the CPU.
However, it is challenging to identify the semantic meaning of raw bytes in memory or on the data bus without having access to additional information \cite{semantic-gap}. For example, reliably determining which memory locations constitute the page table entries or interrupt descriptor tables is quite challenging without looking at the CPU's control registers.

\paragraph{Limitations of Existing Defenses.}
The protections discussed in this section make compromising the OS or planting a rootkit significantly more challenging.
However, as we have highlighted above, these defenses still suffer from a number of limitations and implementation challenges.
Table \ref{table-prot-compare} summarizes these limitations and challenges.

One common theme that can be seen is that a vulnerability in the hypervisor, main kernel, or trusted (isolated) kernel can undo any of the software-based protections. 
And such types of system software are too large to be bug-free -- a point which is evident from the series of attacks we have highlighted above.

The protection we present in this work is aimed at protecting the kernel's integrity even in the presence of driver or kernel vulnerabilities.
%The protection hardware we present in the next section can be configured to disallow any privileged code from modifying security-critical data such as driver signature enforcement configurations -- even after the page permissions have been thoroughly bypassed. 
As we will describe in the next section, the hardware extensions and software modifications required by our protection are minimal, thereby reducing the attack surface.

\section{Neverland: A Hardware Extension for Safeguarding Kernel Integrity}
\label{section-neverland-design}

The previous section discussed how existing OS protections (implemented within a kernel, a hypervisor, or a trusted execution environment) have repeatedly been subjected to attacks as a result of their size and complexity.
In this section, we propose a light-weight, hardware-based protection that enables operating system kernels to protect themselves against tampering -- even in the presence of software vulnerabilities.
The security approach we describe here, dubbed  Neverland, does not rely on hypervisors or additional kernels running in trusted execution environments -- thereby eliminating the complex extra components required by existing protections. 

Neverland's protections work by stripping away the operating system's ability to change values of specific memory locations and configuration registers once the boot process is complete.
Security-critical memory locations, such as security configuration flags, kernel code, and kernel read-only data, do not generally need to be modified once the system is done booting (there are exceptions to this, which we will discuss in Section \ref{restricted-features} ).
Our proposed hardware can be configured at boot time \textbf{to prevent the introduction of any new privileged code} into the system once the kernel is initialized, and \textbf{to disable all writes (at the hardware level) to kernel code and read-only data}. 
Since even the kernel is prevented by the hardware from modifying security-critical code, data, and configurations, software vulnerabilities in the kernel cannot be exploited to alter these memory locations.

We validated the efficacy of these protections by prototyping the hardware extensions in Spike (the official RISC-V emulator) and running a minimally modified version of the Linux kernel on the emulator. 
Our evaluations show that the hardware extensions required by Neverland do not incur any appreciable silicon or energy cost, and that integrating the proposed protections does not incur any performance overhead beyond the 10s - 100s of milliseconds that are required to setup permissions at boot time.

\subsection{Threat Model}

We consider a threat model that is similar to what is typically assumed by OS kernel protection mechanisms. 
We rely on the existence of a secure boot mechanism that ensures the system boots into an untampered kernel.
However, we assume the OS kernel itself could have exploitable software vulnerabilities.
The attacker's aim is to use these vulnerabilities to tamper with kernel code/data, to execute kernel-mode malware, or to hide malicious activity.

\subsection{Design and Implementation Overview}
\label{sec-neverland-overview}
In this section, we present the ingredients that makeup Neverland's protection mechanisms.
We explain why these mechanisms reduce the operating system's attack surface in Section \ref{effectiveness-overview}.

\begin{figure*}
\begin{centering}
	
    \includegraphics[width=0.7\linewidth]{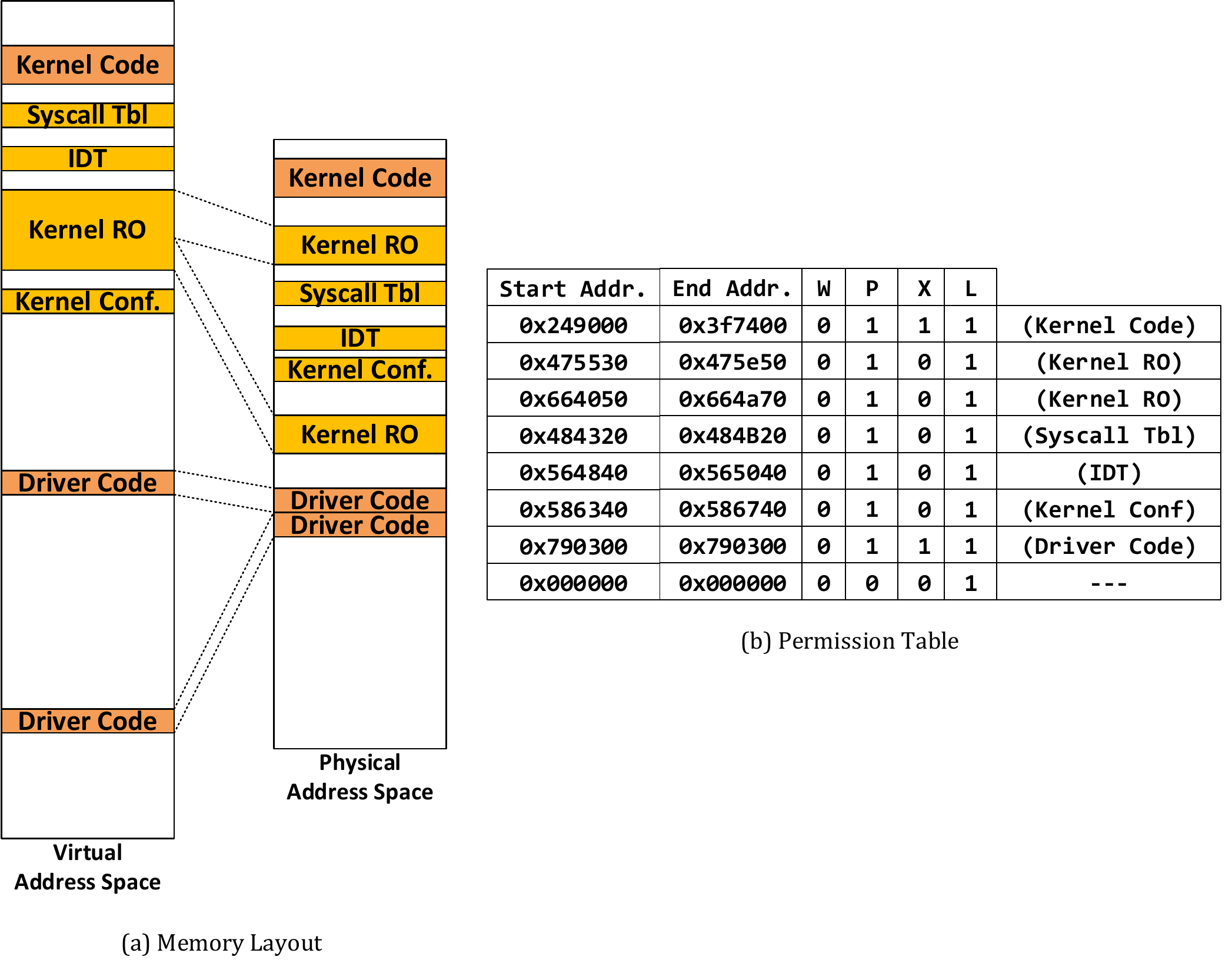}

    \caption[Example Memory Layout and Physical Memory Permissions]{\textbf{Example Memory Layout and Physical Memory Permissions.} Neverland enforces ``write-once'' and non-privileged-code-only restrictions by using a hardware table that stores permissions associated with protected physical address ranges. These permissions are enforced independent of the page permissions maintained by the OS. The permission table shown in (b) illustrates how the memory layout shown in (a) can be protected.  For example, kernel code and read-only data are marked as read-only and locked. Once the locked bit for a permission entry is set to 1, even the operating system (or a malicious privileged code) cannot revert the permission entry until system restart. Hence, marking an initialized memory region as read-only + locked turns it into a ``write-once'' region. Note that virtually contiguous memory locations may not be necessarily be stored in a physically contiguous manner and as a result, we might need multiple permission entries for those regions.}
    \label{fig:phys-va-1}
\end{centering}
\end{figure*}

\paragraph{Write-Once Memory Regions.} 
The kernel code and certain critical data structures (e.g., system call tables, security configuration flags)  do not typically need to change once the system is initialized. 
Even the OS does not need to modify these data structures once they are initialized (there are exceptions to this claim, which we will discuss in Section \ref{restricted-features} ).

Our hardware extensions enable the operating system to disable all \textit{subsequent} writes to selected regions of memory that contain kernel code and security-critical static data -- \textbf{regardless of page permissions or the software privilege level}. 
Furthermore, once these memory regions are marked as read-only, they can be ``locked'' so that even the kernel cannot revert these permissions.
Marking a memory region as locked + read-only once it is initialized effectively turns it into a ``write-once" region.

Note that this write-once capability is entirely independent of the virtual memory system. 
As a result, code/data written in these locked regions will remain protected even if a vulnerability in the kernel is exploited to make read-only pages writable, or to make kernel pages accessible to user space programs.
The only way these ``locked'' and read-only memory regions can be modified or freed is by rebooting the system. This protection approach obviates the need to continuously run an integrity checking thread or virtual machine, or the need to instrument the kernel and redirect page permission management.

\paragraph{Non-Privileged-Code-Only Memory Regions.}
A kernel-mode malware or rootkit will eventually need to execute its own code with escalated privileges.
Neverland's second component is targeted at preventing the introduction of arbitrary privileged code into a system -- even in the presence of vulnerabilities in the OS or drivers.

Neverland enables us to mark certain regions of the \textit{physical} address space as containing kernel-mode code. 
\textbf{All other regions of memory \textit{cannot} contain any kernel or driver code} (i.e., by default, memory regions are only accessible in user-mode).
The CPU cannot fetch code from a physical memory location that is not marked as privileged while it is running in kernel-mode -- and vice-versa. 
Similar to the ``write-once'' regions above, this invariant is enforced by a hardware extension that is independent of the virtual memory system -- making it resilient to attacks even in the presence of operating system bugs.

This feature has two crucial distinctions compared to the Intel SMAP \cite{intel-sdm-3a} and ARM PXN \cite{arm-guide}  features described in the previous section :
\begin{itemize}
\item SMAP and PXN rely on page permission bits and are under the total control of the operating system. As a result, a flaw in the operating system can be exploited to bypass these features \cite{ret2dir,smep-xen-attack,smep-rop}. On the other hand, the non-privileged-code-only memory region proposed here tracks permissions of physical addresses (instead of relying the OS's virtual memory system) and relies on a hardware-managed table that is largely independent of the OS.

\item Once memory locations are categorized as privileged-code and non-privileged-code-only regions (preferably immediately after the kernel finishes booting), the permissions can be locked from subsequent changes -- even by the most privileged software in the system. 

\end{itemize}

This feature, combined with write-once memory regions can be used to ensure all memory regions that contain privileged code cannot be patched, even by the OS or malicious drivers.
Conversely, we can also ensure that writable memory regions can only contain non-privileged code.
This configuration is illustrated in Figure \ref{fig:phys-va-1}.
We can set kernel and driver code as ``locked'' since they do not typically need to change once loaded into memory.
In short, flagging memory regions as locked or non-privileged-code-only makes it extremely hard to introduce new malicious privileged code into the system.

\paragraph{Locking Memory Permission Registers.}
The write-once and the non-privileged-code-only memory restrictions are specified using permission configuration registers (discussed below).
The values of these configuration registers are intended to be set at boot time to minimize the attack surface.
 
After their values have been set, the ability to lock these permission registers against any modification would significantly improve the security of the system.
If these configurations are not locked, vulnerabilities in the OS can be exploited to reset them -- similar to attacks that disable SMAP/SMEP defenses by resetting control registers \cite{smep-xen-attack, smep-rop}.

\paragraph{Locking Configuration Status Registers.}
The CPU determines the location of interrupt descriptor tables (IDTs) and system call dispatchers using values stored in configuration registers. 
An attacker can trick the CPU into executing arbitrary code by overwriting the addresses stored in these registers (provided the attacker has already injected code to be executed into the kernel memory or has bypassed SMEP/PXN).

To provide an additional layer of security, existing kernel integrity mechanisms such as PatchGuard and HyperGuard \cite{windows-7e} monitor configuration registers to detect malicious modifications.
Instead of continuously monitoring changes to these registers in software, ``hardware-locking'' them from subsequent modification provides better security while eliminating the additional monitoring overhead.
We discuss how register locking can be implemented in hardware in Section \ref{hardware-design}.

\paragraph{Permission Tables.} 
The memory protections described above require a lightweight hardware to be integrated into the CPU. 
This additional hardware will maintain a small memory permission table that contains the physical address ranges and their associated permission bits.
Permission bits specify whether a memory range is writable or ``privileged-executable''. A third permission bit is used to ``lock'' the permission entry (as discussed above).  
Figure \ref{fig:phys-va-1}b shows what the permission table's contents would look like for protection setup shown in Figure \ref{fig:phys-va-1}a. 

A number of embedded CPU architectures support a form of memory permission tables. For example, the ARM Memory Protection Unit (MPU) and the RISC-V Physical Memory Protection (PMP) scheme are optional components that can be used to enforce permissions on physical addresses\cite{riscv-privileged-spec, arm-arch-manual}. 
They are a preferred mechanism for implementing a lightweight memory protection scheme in low-resource microcontrollers that do not have virtual memory support \cite{zele-sfi, privilege-overlays}.  These extensions, however, are not typically implemented by CPUs that have virtual memory support.
Neverland uses a variation of such tables for enforcing memory permissions. We leverage immutable hardware tables and registers to provide low-overhead operating system protections.

We will dive into the details of the hardware implementation in the next section.

\paragraph{Permission Initialization and Enforcement.}
A kernel-mode code configures the permission tables immediately after all kernel code and read-only data are loaded into memory.
If a memory region is covered by multiple overlapping permission entries, the most restrictive set of permissions are enforced.
To prevent an attacker from specifying additional privileged code regions, unused permission entries must be locked as shown in the last row of Figure \ref{fig:phys-va-1}b.  
Initializing permissions immediately after secure boot, and locking any unused permission entries takes away the attacker's ability to load arbitrary kernel-mode code. 

\paragraph{What Memory Regions Are Protected?}
Neverland is ideal for protecting security-critical code and data that are loaded or set once (preferably at boot time), and do not need to change once the system is running.
A typical configuration for protecting an OS would be composed of the following set of permissions, which are to be set immediately after the OS is done loading code and data into memory:
\begin{itemize}[leftmargin=*]
    \item \textbf{Core kernel code (text section):} privileged, read-only, executable, locked 
    \item \textbf{Read-only kernel data (RO section):} privileged, read-only, locked
    \item \textbf{System call table:} privileged, read-only, locked
    \item \textbf{Interrupt descriptor table (on x86):} read-only, locked \sfy{verify how new interrupts are registered}
    \item \textbf{Kernel Configuration (e.g., code signing enforcement configurations):} \\privileged, read-only, locked    
    \item \textbf{Driver / loadable kernel module code:}  privileged, read-only, executable, locked

\end{itemize}
Note that the code and data regions listed above are controlled or periodically scanned by a kernel integrity mechanism such as PatchGuard and HyperGuard \cite{windows-7e}. 
The ability to irreversibly disable writes (until system restart) to these regions eliminates the need to monitor these regions using a continuously running software.

On the Linux kernel, the kernel code (text) and read-only data sections are typically stored in a physically contiguous region. So we will only need one permission entry for each. Even a 5-entry permission table would be sufficient to protect the critical components of the core kernel listed above.

On the other hand, loadable kernel modules/drivers are not loaded in a physically contiguous memory region, and as a result, protecting them would require dozens of entries in the permission table. This is undesirable as having a large hardware table incurs high overhead and cost. We will discuss how we tackle this challenge in Section \ref{lkm-packing}. 

\paragraph{Function Pointers.}
As discussed in the previous section, function pointers in the operating system could be overwritten by rootkits to hijack execution of system calls or driver code \cite{hookscout}. 
Function pointers that are typically hooked by rootkits rarely change their values once they are initialized \cite{hooksafe}. 
Neverland could be used to lock such function pointers after they are initialized. Wang et al. have shown such function pointers can be relocated to dedicated read-only pages  \cite{hooksafe}. Their pointer relocation approach can be leveraged by Neverland to prevent malicious function pointer hooking.
However, function pointers that need be updated multiple times while the system is running cannot be locked by Neverland.

\begin{comment}

\subsection{Effectiveness Against Attack Vectors}
\label{effectiveness-overview}

Neverland's physical memory protection can be configured to ensure that
\begin{itemize}
    \item the kernel/driver code cannot be patched without rebooting the system
    \item an attacker cannot run arbitrary code at escalated privileges
    \item security-critical data structures cannot be easily maliciously modified.

\end{itemize}

The protections we present aim to provide identical security guarantees as existing kernel protection features that rely on page access permissions, supervisor mode access/execution prevention \cite{smep, arm_pxn}, and kernel integrity protection/monitoring mechanisms \cite{knox, wind10, virt_based}.
These existing protections, however, have the following limitations:
\begin{itemize}
    \item 
    \item a bug in the operating system can potentially undo any of these protections
\end{itemize}

(use real attack examples to show why OS-independent )

Furthermore, existing kernel integrity mechanisms require that the protected operating system run on top of a hypervisor (or in a trusted execution environment) alongside a second kernel \cite{virt_based}, or need periodically ... All of these

Neverland

- independent of paging - unaffected by vulnerabilities
- make it extremely hard to introduce new code
- makes it hard to patch kernel code
- avoids the need for continuous integrity monitoring
- lower attack surface area

We provide a mode detailed security analysis in Section \ref{sec_eval}
\end{comment}

\subsection{Hardware Requirements}
\label{hardware-design}
%\sfy{keyidea: only need to maintain a permission table, zero pereformance overhead}

As highlighted above, the main additional component that is required is a table that stores and enforces the physical memory permissions. 
%We implemented the hardware table with a series of registers. 
Each permission entry needs to store three pieces of information (Figure \ref{fig:phys-va-1}b):
\begin{itemize}[leftmargin =*]
    \item The physical address range (start and end address) for the memory region
    \item Privileged bit: if this bit is \textit{not} set to 1, the CPU cannot read data or fetch code from this physical memory range while executing in privileged mode. \textbf{By default, all memory regions are non-privileged -- regardless of page permissions.}
    \item Execute bit: the CPU can only fetch kernel-mode code when this bit is set to 1. However, the CPU is still allowed to fetch user-mode code even when this execute bit is set to 0 (i.e., user-mode code execution is enforced by page permission set by the OS).
    \item Write bit: If this bit is set to zero, even the OS cannot write to this memory region, regardless of page permissions.
    \item Locked bit: After this bit is set, the operating system cannot make any modifications to this permission entry. 
    \item Valid bit (optional): this indicates a permission entry has been fully initialized. Writes to the permission registers may not be atomic; hence this bit can be used to make it easier for the software to prevent the hardware from enforcing a permission entry before it is fully initialized. Note that enforcing a partially initialized permission may prevent the operating system from resuming execution. 
\end{itemize} 

The hardware consults permissions in this table on each L1 cache access.
If there are overlapping address ranges in the list of permissions, we enforce the most restrictive ones.

\begin{figure}
    \centering
    \includegraphics[width=0.6\linewidth]{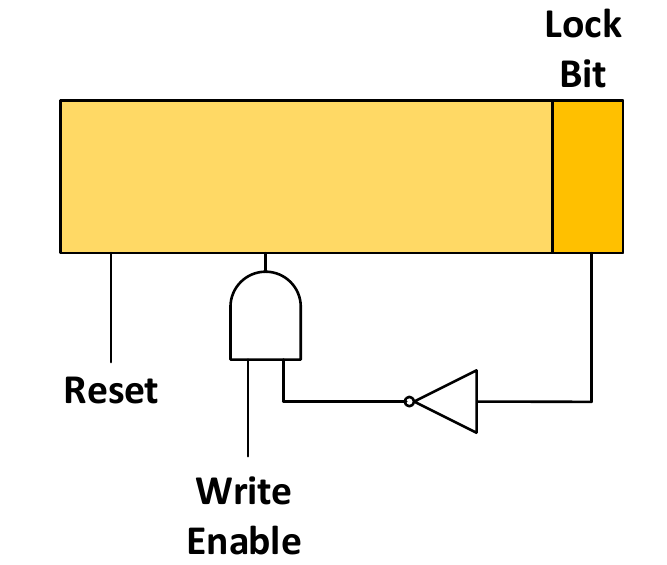}
    \caption[Register Locking]{\textbf{Register Locking.} The lock bit can be used to mask the write enable input. Once locked, the register can only be overwritten after resetting the system.}
    \label{fig:reglock}
\end{figure}

\paragraph{Register Locking.}
As mentioned above, to prevent any additional modifications to a permission entry or a configuration status register, the kernel can set the lock bit associated with that register.
We can use the lock bit to mask the ``write-enable'' input of the register. 
This mechanism is shown in Figure \ref{fig:reglock}.
The only way to reset these registers once they are locked is to reboot the system.

\paragraph{Eliminating Additional Memory Latency.}
\label{zero-overhead}

\begin{figure}
    \centering
    \includegraphics[width=\linewidth]{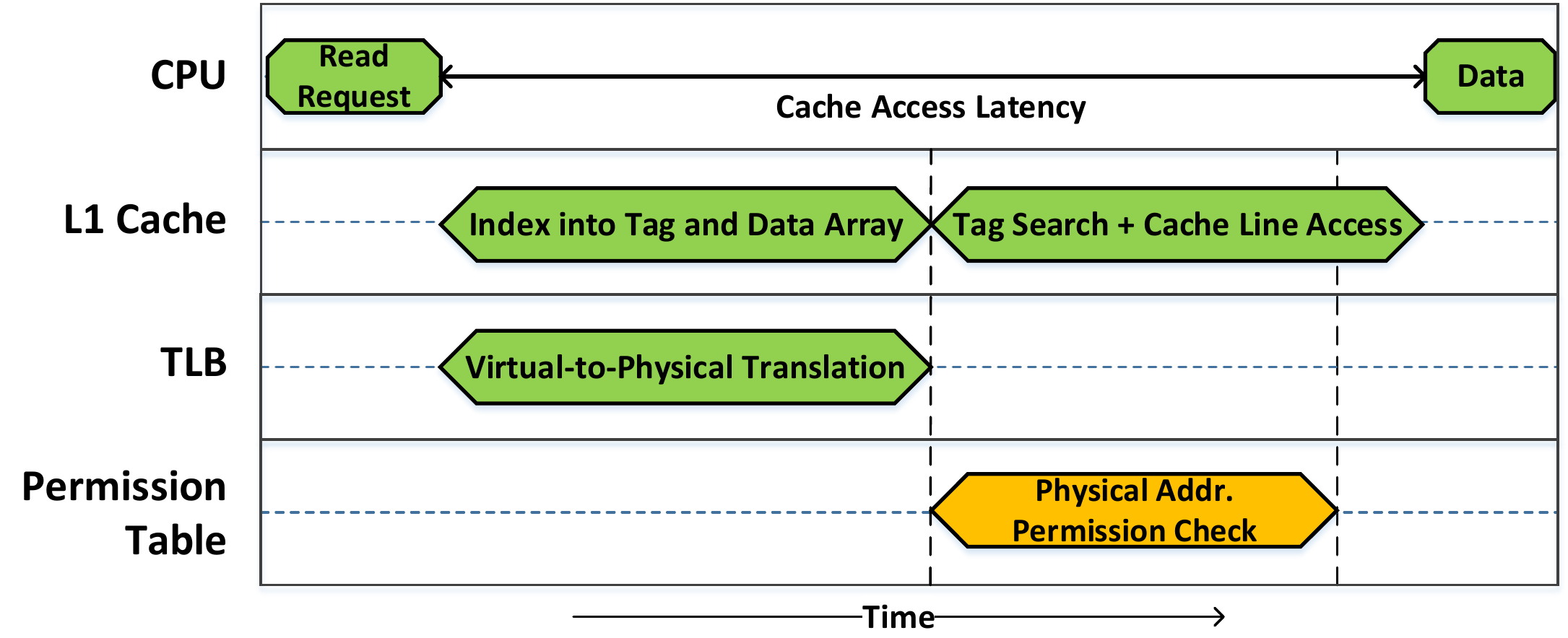}
    \caption[Performing Permission Check in Parallel with Cache Access]{\textbf{Performing Permission Check in Parallel with Cache Access.} After the virtual-to-physical address translation is completed, performing tag search and cache line access requires additional cycles (typically $\sim4$ cycles). This time window can be used to perform permission checks without impacting the cache access latency. The above figure illustrates the case where we have both an L1 cache hit and a TLB hit.}
    \label{fig:cache-access}
\end{figure}

The kernel integrity enforcement scheme we are proposing requires a permission check operation on every data and instruction fetch. 
Therefore, it is essential that the permission lookup process does not stall the CPU lest it will incur overheads on every cache read operation.

\paragraph{Keeping Permission Tables Small.}
To do quick permission lookups, we need to keep the hardware table small. 
Searching through a hardware maintained table will be slower and more expensive (in terms of energy consumption and silicon cost) as the size of the table grows.

If the security-critical kernel memory regions are spread across numerous physically non-contiguous memory regions, then we will need a large permission table to protect them.
To protect the operating system using a small, high-speed permission table, we make minor modifications to the kernel to store the regions that need to be protected by Neverland within a limited number of physically contiguous memory locations. These software modifications are discussed later in Section \ref{lkm-packing}.
In our evaluations, an 8-entry table was quite sufficient to protect all critical memory regions in the Linux kernel. Our evaluations also show that such small tables can be integrated into high-frequency CPUs without impacting cycle-time, and can be searched within 3 CPU cycles (Section \ref{sec-hardware-overhead}). 

\paragraph{Parallelizing Permission Lookups.}
The permission checks can be timed in a manner that ensures the extra cycles that are required to search the table do not stall the CPU. 
Figure \ref{fig:cache-access} illustrates the operations involved in a cache access. 
Initially, the CPU issues a virtual address it wants to access. 
This address is translated to a physical address using the translation lookaside buffer (TLB).
On high-performance CPUs, portions of virtual address bits are used to initiate cache access -- in parallel with the address translation process (more formally, the virtual address is used to index into the tag and data arrays). 
Once the address translation process is done, the physical address is used to search the tags and select the appropriate data array. 

As it can be seen in the timing diagram in Figure \ref{fig:cache-access}, once the physical address is returned by the TLB, there is time window at the end in which the cache performs tag comparisons and reads the appropriate cache line.
The physical memory permissions can be read in parallel with this last operation -- without stalling the CPU.
On today's high-performance CPUs, the tag-compare and data-access operations in the L1 cache take $\sim4$ CPU cycles \cite{hennessy-patterson}. The results we present in Section \ref{sec-neverland-eval} show that this 4-cycle time window is enough to perform physical address permission lookups in parallel.
As a result, Neverland's protections do not incur any performance overheads.

On CPUs without virtually indexed caches (i.e., that do not overlap cache and TLB lookup ), it is relatively more straightforward to do permission lookup in parallel with cache access as the tag-search and data-array-select operations take even longer.

\begin{comment}
    
\subsection{Leveraging Existing Hardware IP}
\sfy{is this necessary? move to evaluation?}
\key{s}

As mentioned earlier, both academic and commercial designs have implemented different variations of memory permission tables for various purposes.
These existing hardware designs can be modified and adopted by CPU designers to enable the protections discussed in this work. 
To illustrate how existing hardware IP can be used. 

\begin{itemize}

\item \textbf{Intel CPUs:} Recent CPUs from Intel implement a hardware extension, known as SGX (), that can be used to shield application modules from untrusted or compromised operating systems. SGX's underlying hardware mechanism maintains a hardware permission table ...
\item \textbf{ARM and RISC-V:} Software running on low-resource or real-time embedded devices do not typically support virtual memory. As a result, they cannot benefit from page based permissions. To mitigate this shortcoming, ARM and RISC-V devices provide a small permission table that can be used to set read/write/execute permissions of different memory regions. These tables can be extended to provide the full set of functionalities required  by (e.g. ) 
\item \textbf{Debug Registers:} Most CPU architectures today provide debug registers that can be configured to monitor read, write, or instruction fetches from different addresses. The hardware required for Neverland can be derived from these ...

\end{itemize}
\end{comment}

\subsection{Supporting Loadable Kernel Modules}
\label{lkm-packing}
%\key{loaded at arbitrary, physically non-contiguous memory addresses.... makes it for hardware}

Modern operating systems need to support a vast array of platforms and peripherals. 
Compiling the kernel with all the drivers or OS extensions that will ever be required would result in an extremely bloated operating system.

To that end, operating systems typically allow drivers to be implemented as separate kernel modules that can be dynamically linked to the kernel as necessary. 
The core kernel loads all the modules that are required to run the OS on the target hardware configuration.
On a typical Linux machine, for example, there could be 10s or 100s of kernel modules that need to be loaded into the system (on a laptop we edited this section on, for example, a total of 171 kernel modules were automatically loaded at boot time).

\begin{figure}
    \centering
    \includegraphics[width=0.7\linewidth]{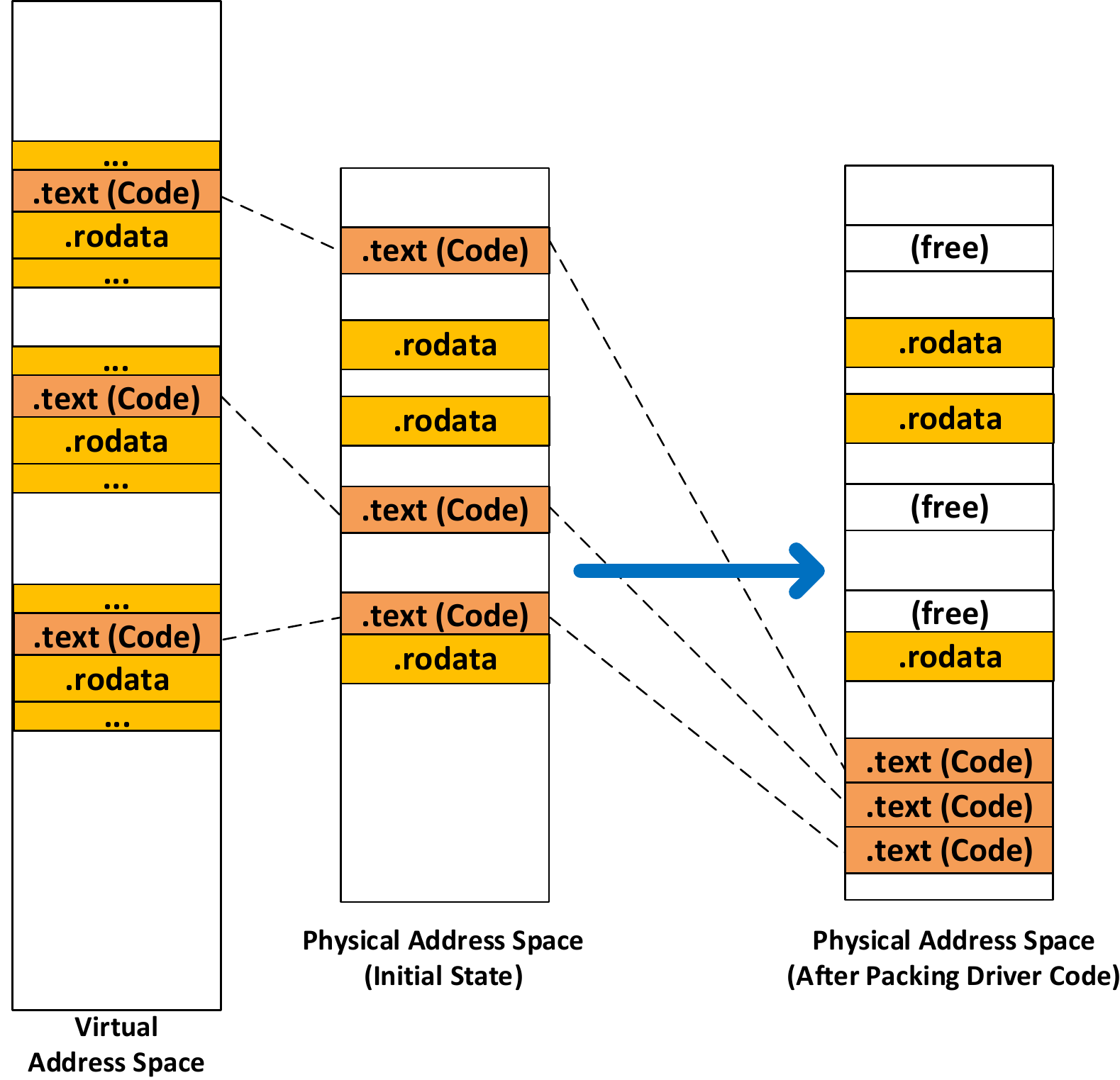}
    \caption[``Defragmenting'' Kernel Module Code]{{\textbf{``Defragmenting'' Kernel Module Code.}} The code sections of loadable kernel modules are placed in numerous physically non-contiguous memory regions. As a result, protecting them directly would require numerous entries in the permission table. To avoid the need for large hardware tables, we re-arrange the physical memory layout and place all the kernel modules' code sections in a physically contiguous region. This enables us to protect all code sections using a single entry in the permission table. The algorithm we present in Section \ref{lkm-packing} can be used to ``defragment'' any kernel-mode code/data that needs to be protected by Neverland.}
    \label{fig:code-packing}
\end{figure}

Loadable kernel modules present a challenge to Neverland's memory protection mechanisms.
In addition to protecting the core kernel, Neverland needs to protect the text (code) section and security-critical data of all loaded kernel modules -- \textbf{which are going to be loaded at arbitrary, physically non-contiguous memory addresses.} 
This fragmentation is illustrated in Figure \ref{fig:code-packing}. 
Each of the potentially 100s of modules loaded into the system could have code and data that span multiple virtual pages that are \textit{not} physically contiguous.

If Neverland ends up maintaining a large table with 100s of entries, then the permission lookup operation will be slow -- which incurs performance overheads.
Furthermore, having large permission tables will be impractical in CPUs used by mobile and embedded devices -- due to their cost and energy constraints.

\paragraph{Rearranging Physical Memory Layout.} 
To enable Neverland to support any number of kernel modules in a scalable manner, we need to modify the kernel module loader logic. 
In this work, we modified the loadable kernel loader in the Linux kernel so that all text (code) sections from loadable kernel modules will be laid out on a physically contiguous memory region.
Although the modifications and results we present in this work are in the context of the Linux kernel, the algorithm presented here can be adopted into any other OS. Furthermore, the same algorithm can be used to contiguously store memory allocated by any kernel subsystem.

The steps taken by our algorithm for laying out kernel modules is illustrated in Figure \ref{fig:code-packing} and is summarized below:

%\noindent\rule{0.5\textwidth}{0.4pt}
\begin{itemize}
        \item \textbf{Step 1: \textit{Allocate a large physically contiguous region}} before any kernel module is loaded (or at boot time). The size of this contiguous region can be configured as a kernel parameter, and can be made arbitrarily large since any unused portions will be freed at the end.
        
    \item \textbf{Step 2: \textit{for each loadable kernel module}}:
\begin{itemize}
    \item \textbf{Step 2.1: \textit{Load kernel modules using the regular flow.}} We do not need to modify the core logic for linking and loading kernel modules.
    
    \item \textbf{Step 2.2: \textit{Copy the content of the pages we want to protect}} to the physically contiguous region.
    
    \item \textbf{Step 2.3: \textit{Overwrite page table entries (PTEs)}} for the copied pages so that the PTEs point to a physical address in the physically contiguous region.
    
    \item \textbf{Step 2.4: \textit{Flush TLB}} to ensure the hardware will see the modifications to the kernel's PTEs.
    
    \item \textbf{Step 2.5: \textit{Free the old physical page frames.}}
\end{itemize}

\item \textbf{Step 3: \textit{Free any unused memory from the physically contiguous pool}}, after all modules have been loaded.

\end{itemize}
%\noindent\rule{0.5\textwidth}{0.4pt}

At the end of the above operations, the \textit{virtual address} space remains unchanged, and as a result, the rest of the kernel does not need to be modified to support these changes.

As we will detail in the next section, all loadable kernel modules and drivers will have to be loaded at boot time for the system to fully benefit from Neverland's protections.

\subsection{Restrictions on Kernel Features}
\label{restricted-features}
To benefit from the maximum security guarantees afforded by Neverland, the kernel must mark all privileged code read-only, and also lock the permission table entries. 
Such tight restrictions can go against some legitimate system features.
In this section, we identify four such features and explain how they are affected by Neverland.

\paragraph{Runtime Kernel Module and Driver Loading.}
Driver software and kernel modules are typically loaded into a system at boot time.
Most operating systems, however, allow drivers or kernel modules to be loaded at any arbitrary time after the system is done booting.

To allow drivers to be loaded anytime while the OS is running, Neverland's ``non-privileged-code-only'' restrictions on memory regions must be disabled. 
This weakens the security guarantees provided by Neverland.
Therefore, to ensure maximum security, it is desirable to require all necessary kernel modules to be loaded at boot time, and then lock all privileged code regions from additional modification.
With this restriction, the system must be restarted when a driver needs to be updated or installed.

Requiring a restart on OS or driver update will only be a slight inconvenience for regular users, especially on mobile and desktop devices, as all necessary drivers and kernel modules are typically loaded at boot time; and restarting the system on OS and driver updates is already a common practice.
On Android systems, for example, it is already a recommended practice to load all required kernel modules in one pass when the kernel is initialized (for improved performance) \cite{android-lkm}.

%One notable exception is a device driver development and testing machine -- which might require repeatedly reloading driver code. Neverland's features need to be disabled on such machines, just like driver signing is typically disabled on development and testing machines.   

\paragraph{Self-Modifying Code and Just-in-Time Compilation.}
Most modern CPUs allow pages to be marked non-executable. A typical page that contains code is marked read-only and executable, whereas all other pages are set to be non-executable.

However, supporting self-modifying code or just-in-time compilers requires code pages to be writable at least temporarily -- which can potentially expose the system to code injection attacks.
%To protect a system against code-injection attacks,  Apple's iOS, for example, disallows pages to be both writable and executable at any given time -- except by special software that is signed using a unique key \cite{xx}.
To protect a system from injection of kernel-mode code, a Windows 10 system protected by Microsoft's Device Guard does not allow drivers to dynamically allocate executable pages \cite{windows-7e}.
Neverland's protections can be enabled on such restrictive systems without requiring significant modifications.

Modern Linux distributions, however, support an in-kernel just-in-time compiler, known as eBPF (extended BSD Packet Filter), that enables user space programs to run sandboxed bytecode in kernel-mode. This feature is an extension of the original BSD Packet Filter system that enables high-speed packet filtering by avoiding expensive copies and context switches between user and kernel-mode \cite{bpf}.

In-kernel JITs such as eBPF, cannot properly function with the strict configuration presented in the previous sections. 
However, some of these restrictions can be relaxed to allow in-kernel JITs. 
Instead of making \textit{all} privileged-code regions unwritable, Neverland can be configured to allow writes to a fixed, pre-allocated executable memory region that is used by the in-kernel JITs \textit{only}.
All other privileged code (core kernel and driver code) will still be unwritable until a system reset.

It is conceivable that an attacker could find a bug that would allow code-injection into the fixed memory region that is explicitly pre-allocated for the in-kernel JIT. 
However, even with only portions of the privileged code regions marked as write-once, Neverland would still make privilege escalation attacks significantly harder compared to a system that only relies on page permissions.

\paragraph{Live Kernel Patching.}
To avoid the need to restart a server on an OS update, some operating systems support live kernel patching (also called hot patching) -- whereby parts of the kernel code can be overwritten without restarting the system \cite{live-patching, windows-7e}. 
The protections introduced in this paper, by design, disallow modification of kernel code once the boot process is done. 
This effectively makes live kernel patching illegal.
\sfy{it should be noted ...}

On a cloud deployment, the hypervisor and/or host OS can be protected using Neverland, and still allow guest operating systems to be live patched. 
Patching a Neverland protected kernel that is directly running on the hardware, however, will require a system reboot.

Kernel live patching is \textit{not} a critical feature on desktop and mobile systems, and as such Neverland's protections will not introduce significant usability issues.

\paragraph{In Summary...}
Neverland's full kernel protections can be readily adopted on mobile, embedded, and desktop platforms as these machines do not need to load new drivers frequently, do not need to rely on in-kernel JITs, and can usually be restarted when the OS or drivers need to be updated.

On certain server deployments, however, enabling the full set of defenses we presented would require either disabling features such as live kernel patching and privileged JITs, or relaxing some of the restrictions (e.g., allowing the OS to update some privileged code regions).

\section{Security Analysis}
\label{effectiveness-overview}

As discussed in Section \ref{background_defenses}, numerous defenses have been proposed to harden operating systems against attacks.
Among these defenses, Neverland requires secure boot and driver signature verification to ensure that a tampered kernel or a malicious driver is not loaded before the memory permission tables are initialized.
Neverland cannot provide reliable protection if the attacker compromises the secure boot chain before the permission table is initialized.
%On the other hand, it provides better security guarantees while incurring lower overhead compared to existing kernel integrity monitoring schemes (Section \ref{sec-neverland-overview}).

\begin{table*}[t]

%\begin{centering}

\caption[Effectiveness of Neverland's Protections]{\textbf{Effectiveness of Neverland's Protections.} Neverland's hardware controlled memory permissions are effective at a protecting against most OS attack vectors. Code pointer hooking and direct kernel object manipulation (via code-reuse attacks), however, are not thwarted by Neverland.}

\begin{tabularx}{\textwidth}{X|X}
	\hline 
	\textbf{Attacks} (see Section \ref{sec-kernel-attacks}) & \textbf{Protection by Neverland} \\ 
	\hline 
	\hline
	Syscall Table (SSDT) and Interrupt Descriptor Table (IDT) Hooking \cite{subverting-windows} & Full Protection: tables unwritable after boot \\ 
	\hline 
	Configuration (Model Specific) Register Hooking \cite{subverting-windows} & Full Protection: registers that hold system call and interrupt entry points are locked after boot \\ 
	\hline 
	Runtime Code Patching \cite{subverting-windows} & Full Protection: text sections are locked; no privileged code can exist outside of the locked area \\ 
	\hline 
	Code Pointer Hooking \cite{hookscout} & Limited Protection: kernel pointers that need to change overtime cannot be locked; an attacker can hook these pointers, but they can only point to existing kernel code (i.e., cannot point to new attacker injected code).
	\\ 
	\hline 
	Direct Kernel Object Manipulation (DKOM) \cite{subverting-windows} & Limited Protection: objects that need to be updated over time cannot be locked; an attacker can use code-reuse attacks or exploit memory management bugs to overwrite objects.
	\\ 
	\hline 
	Malicious Drivers \cite{subverting-windows} & Full Protection: no new privileged code can be loaded after boot
	\\ 
	\hline 
	Privilege Escalation (User space to Kernel space ) (\S \ref{sec-kernel-attacks}) & Full Protection: no code outside the locked region can execute with kernel privileges
	\\ 
	\hline
	 
\end{tabularx} 

%\end{centering}

\label{table-attack-vectors}

\end{table*}

Table \ref{table-attack-vectors} highlights how Neverland helps protect against existing attack vectors and stealth techniques.
It can be seen from the table that Neverland's hardware controlled memory permissions are effective at thwarting most attack vectors.
The two exceptions are code pointer hooking and direct kernel object manipulation. 
Indefinitely locking arbitrary code pointers and kernel objects from modifications is not feasible since the operating system might need to update them as the state of the system changes.
As a result, an attacker could still hook code pointers or leverage code-reuse techniques such as return-oriented-programming \cite{rop} to maliciously modify kernel objects on a system protected by Neverland.

Even if Neverland cannot be used to directly protect dynamic kernel objects and pointers, the fact that it can totally disable any unauthorized privileged execution makes it hard to meaningfully manipulate these unprotected pointers and data structures. The attacker will need to purely rely on code-reuse attacks to modify desired memory regions. Furthermore, after hooking code pointers, the attacker cannot redirect them to new privileged code, but only to existing kernel/driver code. 
Operating systems can also use Neverland to ``lock'' kernel configuration flags, such as driver signature enforcement configurations -- which are data structures typically targeted by kernel exploits  \cite{pegasus, uroburos}.
Hence, the protections we have presented in this work significantly limit what the attacker can achieve by way of kernel exploits.

\begin{comment}

\subsection{Protection Against Unauthorized Privileged Execution}

\subsection{Protection Against Stealth Techniques}

The most insidious malware not only compromise the integrity of the OS to execute privileged code, but also employ techniques to avoid detection. 
These techniques are used to hide running processes, network connections, or even entire file systems so that the user or any other programs cannot detect the existence of the rootkit.

Such stealth is normally achieved using Direct Kernel Object Manipulation (DKOM), hooking, or code patching. As discussed above, the write-once memory approach prevents the possibility of code patching. However

\paragraph{Direct Kernel Object Manipulation:} direct kernel manipulation 

\paragraph{Driver Signing} 
Requiring drivers to be signed using 

\paragraph{Supervisor Mode Execution Prevention} ..

\paragraph{Kernel Integrity Monitoring} ..

\paragraph{Virtualization Based Security} ...
\end{comment}

\section{Evaluation}
\label{sec-neverland-eval}
We evaluated the effectiveness and practicality of Neverland by adding the defenses presented above to a RISC-V system. We extended Spike, the official RISC-V ISA emulator \cite{riscv-software}, with the proposed hardware extensions. We modified the Linux kernel to i) initialize the permission tables, ii) lock the configuration registers which store the addresses of the trap handlers (mtvec and stvec registers on RISC-V), and iii) defragment loadable kernel modules.
RISC-V already has an (optional) ISA extension for physical memory protection \cite{riscv-privileged-spec} -- which is an approach typically used as a replacement to page-based permissions on low-resource microcontrollers \cite{zele-sfi, privilege-overlays}. Hence, we use those instructions to write to the permission registers. On other architectures, the registers in the permission table could be programmed through a memory-mapped I/O.

We were able to boot a fully functional 64-bit Linux-based system with BusyBox 1.26.2 utilities \cite{busybox} on the Spike emulator with these protections in place. 
This validates that a stock kernel (with minor modifications) can run with portions of its memory completely locked-down by the hardware.
We initialize the permission table once the kernel is fully initialized and read-only page permissions are set up by the kernel.
In the remainder of this section, we evaluate the overheads associated with these changes.

\subsection{Hardware Overhead}
\label{sec-hardware-overhead}
The only additional hardware that is necessary for locking-down memory locations is the permission table and the permission check hardware.

To measure permission lookup latency and the hardware cost, we implemented a permission table and lookup logic in RTL and used the Synopsis Design Compiler to synthesize the designs to the IBM 45nm silicon-on-insulator (SOI) technology library.

\paragraph{Speed.}
We were able to clock our synthesized permission table at frequencies as high as 3.83 GHz on a 45nm library.
For comparison, the highest peak frequency available on a 45nm Intel CPU is 3.73GHz \cite{cpudb}. 
This confirms permission tables can be integrated even in high-frequency CPUs without affecting speed.
45nm is a relatively older technology node, but this comparison holds true on newer silicon technology as well since both the permission table and rest of the CPU scale in a similar manner.

Furthermore, using this design, it is possible to perform a lookup in a 16-entry permission table in just 3 cycles. 
And as described in Section \ref{zero-overhead}, this 3 cycle table lookup can be performed in parallel with the tag comparison and data array access -- without incurring any performance overhead.

%and the design can be clocked at higher frequencies if implemented on newer nodes. 
%For example, a 16nm process can conservatively have $\sim1.25x$ higher clock rate compared to a 45nm node -- enabling us to clock the design at $\sim4.6GHz$ without any modifications to the design.
%To sum up, the permission table can be searched through in just 3 cycles -- even on high-frequency CPUs.

%Table xx shows the number of cycles required for searching permission tables of different sizes. 
\begin{table*}[t]
    \begin{center}
    	
    \caption[Power and Area Overhead of Permission Tables]{\textbf{Power and Area Overhead of Permission Tables.} Overheads normalized with respect to a 40nm dual-core ARM Cortex-A9 (a CPU targeted at low-power, cost-sensitive embedded systems), a single-core 45nm RISC-V BOOM CPU with no L2 cache, and a 45nm quad-core desktop-class Intel Core i5-760 CPU. The estimates for the mobile CPUs are pessimistic as the permission tables are not synthesized to a power-optimized process node.}

        \begin{tabular}{|c|c|c|c|||c|c|c|}
            \hline 
            No. of Entries & \multicolumn{3}{c|||}{Normalized Area Overhead} & \multicolumn{3}{c|}{Normalized Power Overhead}  \\ 
            \cline{2-7} 
        %    Entries & Normalized to & Normalized to & Normalized to & Normalized to \\ 
            (per core) & BOOM &  Cortex-A9 & Core i5      &   BOOM        & Cortex-A9  & Core i5 \\
            \hline \hline \hline
            4 x 2    &  0.67\% &    0.59\% & 0.016\%    &  4.29\%    & 1.79\%  &  0.07\% \\
            \hline
            8 x 2    &  1.26\% &    1.11\% & 0.032\%    &  7.69\%    & 3.20\%  &  0.13\% \\
            \hline
            16 x 2    &  2.39\% &    2.10\%    & 0.062\%   & 14.18\%    & 5.91\%  &  0.23\% \\
            \hline
            
        \end{tabular}
    \end{center}
    
    \label{table-area-power}
\end{table*}

\paragraph{Hardware Area and Energy.}
Table \ref{table-area-power} lists the estimated power and area overheads of permission tables synthesized with different sizes. We present the overheads normalized with respect to i) a 40nm, dual-core ARM Cortex-A9 (a CPU targeted at low-power, cost-sensitive embedded systems \cite{cortexa9}), ii) a single-core 45nm, RISC-V BOOM with no L2 cache (a CPU that targets similar device classes as the Cortex-A9), and iii) a 45nm, Intel Core i5 (a desktop-class CPU).
The power and area of the two baseline embedded/mobile CPUs is based on the data from \cite{boom}, while the corresponding estimates for the permission tables are based on a 45nm synthesis result at a target frequency of 1.5GHz.
The baseline power and die size for the desktop-class CPU was acquired from the Core i5-760 datasheet, while its corresponding permission table is synthesized with a target frequency of 3.8GHz.
The estimates account for 2 permission tables \textit{per core} -- one for the I-cache and a second one for the D-cache.

The results show that adding 8-entry permission tables even on low-end embedded CPUs such as the Cortex-A9 incurs minimal overhead -- 1.11\% area overhead and 3.2\% power overhead.
These estimates are pessimistic as the permission tables are not synthesized to energy optimized process.
These overheads will be even lower on CPUs that do not merely target low-end devices.
On smaller CPUs such as the BOOM or Cortex-A5, adding more than 4-entry permission tables (without power optimizations) would result in a sizable power overhead.

\subsection{Performance Overhead}
The evaluation above indicates that permission lookups, which are performed on each L1 cache access will not degrade system performance.
In addition to the hardware extension, however, we made two additions to the kernel code, namely packing the text section of loadable kernel modules and setting up the permission table at boot time.
We evaluate the performance impact of these kernel modifications on boot latency below. 
Note that these two modifications only impact the boot latency as they only need to be executed once on system startup.

\subsubsection{Kernel Module Loading}
At the time of this project, the upstream RISC-V Linux port (v. 4.14) did not yet fully support dynamically loadable kernel modules. 
Hence, we validated our kernel module packing scheme and characterized its performance on an x86\_64 system. 
We implemented the algorithm described in Section \ref{lkm-packing} in version 4.14 of the kernel, and run it on a KVM virtual machine. \sfy{TODO: more about the machine, averaged 4 measurements}

\begin{table}

    \begin{center}
		
		\caption[Module Load Latency Overhead Incurred by Defragmentation]{\textbf{Module Load Latency Overhead Incurred by Defragmentation.}}
        
        \begin{tabular}{|c|c|c|c|}
            \hline 
            &       \multicolumn{3}{c|}{\textbf{Module Load Time}}   \\
            \cline{2-4}
            \textbf{Code Size} & \textbf{Baseline} & \textbf{Packed Text}  & \textbf{Overhead} \\ 
            &        &                       \textbf{Section}    & \\
            \hline 
            \hline
            128KB & 4.86 ms & 5.89 ms & 1.03 ms \\
            256KB & 5.15 ms & 6.27 ms & 1.13 ms \\
            512KB & 5.91 ms & 7.28 ms & 1.37 ms \\
            1MB   & 7.32 ms & 8.67 ms & 1.35 ms \\
            \hline 
        \end{tabular} 
        
    \end{center}

    \label{table-module-time}
\end{table}

To measure the additional latency incurred on kernel module loading as a result of moving code pages, we compiled and loaded kernel modules with different code sizes. The results are presented in Table \ref{table-module-time}. The results demonstrate that relocating even 100s of KBs of code pages incurs $< 1.5ms$ overhead. 

\subsubsection{Initializing Permission Registers}
After the boot process is completed (including kernel module packing), we need to write the appropriate permissions to the hardware table. When booting the Linux kernel on a single emulated core in the Spike emulator, writing the addresses and permissions to the hardware table only introduced an additional $270\mu s$ latency on the boot process.
We expect the latency to be even lower on real hardware that does not have the emulation overhead. 
Note that the initialization needs to be done on every core on a multicore system -- similar to any other configuration status registers.

\section{Conclusion}

In this work, we presented a hardware mechanism for thwarting kernel-mode malware and rootkits. 
Our approach prevents runtime modification of the kernel's critical code/data by locking them at the hardware level once the boot process is complete.
Creating a full-fledged kernel-mode malware or rootkit by relying entirely on code-reuse attacks and kernel object manipulation is challenging.
Hence, taking away the ability to execute malicious kernel-mode code or to ability modify static kernel data significantly limits the attacker's facilities. 

\section*{Acknowledgments}
This work was supported by DARPA under Contract HR0011-18-C-0019. Any opinions, findings andconclusions or recommendations expressed in this materialare those of the authors and do not necessarily reflect theviews of DARPA.

\printbibliography

\end{document}